\documentclass[conference]{IEEEtran}
\IEEEoverridecommandlockouts

\usepackage[noadjust]{cite}

\usepackage{amsmath,amssymb,amsfonts}
\usepackage{algorithmic}
\usepackage{graphicx}
\usepackage{textcomp}
\usepackage{xcolor}
\usepackage{amsmath}
\usepackage{amssymb}
\usepackage{dsfont}
\usepackage{url}
\usepackage{multirow}
\usepackage{pdfpages}

\def\BibTeX{{\rm B\kern-.05em{\sc i\kern-.025em b}\kern-.08em
    T\kern-.1667em\lower.7ex\hbox{E}\kern-.125emX}}

\newcommand{\bD}{\mathbb{D}}
\newcommand{\bE}{\mathbb{E}}
\newcommand{\bR}{\mathbb{R}}

\newcommand{\cL}{\mathcal{L}}
\newcommand{\cO}{\mathcal{O}}

\newcommand{\bn}{[\![1,n]\!]}
 
\begin{document}
\title{Multiple Choice Learning for Efficient Speech Separation with Many Speakers}

\author{\IEEEauthorblockN{David Perera$^1$\IEEEauthorrefmark{1}, François Derrida$^1$\IEEEauthorrefmark{1}, Théo Mariotte\IEEEauthorrefmark{2},  Gaël Richard\IEEEauthorrefmark{1}, Slim Essid\IEEEauthorrefmark{1}}
\IEEEauthorblockA{\IEEEauthorrefmark{1}LTCI, Télécom Paris, Institut polytechnique de Paris, France}
\IEEEauthorblockA{\IEEEauthorrefmark{2}LIUM, Le Mans Université, Le Mans, France}
}

\maketitle

\begin{abstract}
Training speech separation models in the supervised setting raises a permutation problem: finding the best assignation between the model predictions and the ground truth separated signals. This inherently ambiguous task is customarily solved using Permutation Invariant Training (PIT). In this article, we instead consider using the Multiple Choice Learning (MCL) framework, which was originally introduced to tackle ambiguous tasks. We demonstrate experimentally on the popular WSJ0-mix and LibriMix benchmarks that MCL matches the performances of PIT, while being computationally advantageous. This opens the door to a promising research direction, as MCL can be naturally extended to handle a variable number of speakers, or to tackle speech separation in the unsupervised setting.
\end{abstract}


\def\thefootnote{1 }\footnotetext{Equal contribution.}

\section{Introduction}\label{sec:introduction}

Speech separation is the task of isolating concurrent speech sources from a mixture in which they are simultaneously active. 
This task has many applications in speech processing, including Automatic Speech Recognition \cite{marti2012automatic,li2021espnet}, Speaker Diarization \cite{von2019all, boeddeker2024ts}, or singing voice extraction from music~\cite{huang2012singing}.
Various scenarios of practical interest involve a large number of simultaneously active speakers (cochlear implants~\cite{loizou2009speech} or human-robot interactions~\cite{deleforge2012cocktail}). 
While previous works mainly focus on the few-speaker setting (up to five active speakers \cite{horiguchi2020end,lee2024boosting}), speech separation with many speakers \cite{ephrat2018} is still a critical area for improvement \cite{lutati_sepit_2023}.

The state-of-the-art approach for speech separation consists in training a deep neural network 
in a supervised setting, where the ground truth for the individual sources is known~\cite{wang2018supervised}.
In this setting, the separation model takes as input a single-channel mixture, and predicts a fixed number $n$ of individual tracks.
These predictions are compared to the ground truth speech sources in a pairwise fashion using an audio-reconstruction metric, such as the scale-invariant Signal-to-Distortion Ratio (SI-SDR) \cite{le2019sdr}. 

This pairwise association is ambiguous, and raises a label permutation issue: finding the optimal matching between the set of predictions from the model and the set of ground truth separated signals. 
Early attempts include speaker-dependent methods~\cite{huang2015joint, weng2015deep}, deep clustering \cite{hershey2016deep} and deep attractors~\cite{chen2017deep}, which rely on speaker statistics or clustering techniques to accurately associate predictions to target.
However, these heuristics lead to suboptimal assignations. PIT \cite{yu2017permutation} solves this issue by proposing a much simpler but computationally expensive framework: optimizing, among all possible prediction--target matching, the one that minimizes the global separation error.

A naive implementation of PIT has an intractable complexity of $\cO(n!)$. This has been a bottleneck when training speech separation systems for a large number of speakers. However, PIT has been reformulated as a perfect matching problem in a bipartite graph, thus improving its complexity to $\cO(n^3)$ by using the Hungarian algorithm \cite{dovrat_many-speakers_2021, edmonds1972theoretical}. 
Recognizing the assignation task as an instance of the optimal transport problem, this complexity has been slightly reduced to $\cO(n^2/\varepsilon)$ with the Sinkhorn algorithm (SinkPIT), at the cost of an approximation controlled by a parameter $\varepsilon$~\cite{tachibana_towards_nodate, cuturi2013sinkhorn}. 

Recently, the authors of \cite{perera2024annealed} have proposed to use the Multiple Choice Learning (MCL) framework \cite{guzman2012multiple,lee2016stochastic}, which has~$\cO(n^2)$ complexity. Unlike PIT, this method is not guaranteed to provide the optimal prediction--target assignation. Instead, MCL is designed to tackle ambiguous tasks, such as multi-modal trajectory forecasting \cite{wang2020dsmcl}, where the relation between input and target is non-deterministic, and multiple predictions should be provided to capture the resulting uncertainty. 
Reframing speech separation as an ambiguous task,
the authors of \cite{perera2024annealed} propose to use the multiple predictions provided by MCL as estimations of the separated speech signals.
They demonstrated empirically that this approach was effective for 2-speaker and 3-speaker mixtures. 
The purpose of this paper is to extend this result to the many speakers regime.
More specifically, we make the following contributions:

\begin{itemize}
\item  \textbf{We establish MCL as a viable speech separation method}, even in the challenging many speakers setting. 
\item  \textbf{We compare MCL, Hungarian-PIT and SinkPIT} in terms of training time and separation performance. 
\item  \textbf{We introduce AUC-SDR}, a metric evaluating the consistency of reconstruction quality across separated sources, which becomes especially relevant when separating many speakers.
\end{itemize}

This paper is organized as follows. Section \ref{sec:method} describes the compared training frameworks, Section \ref{sec:setup} details our experimental setup, and Section \ref{sec:results} discusses our results.

\section{Method}\label{sec:method}

\subsection{Training framework}

Formally, let $y=(y_1, \ldots, y_n)\in\bR^{l\times n}$ denote audio signals from $n$ individual speakers, with $l$ their common length measured in time frames. The task of speech separation consists in providing an estimate $\hat{y}=(\hat{y}_1, \ldots, \hat{y}_n)\in\bR^{l\times n}$ of the isolated speech signals from a mixture $x\in\bR^l$ using a neural network $f_\theta$ with parameters indexed by $\theta$. 

The model $f_\theta$ is customarily trained using gradient descent by optimizing the expected value $\bE_{(x,y)\sim\bD}\left[\cL(y, f_\theta(x))\right]$ taken by a separation loss $\cL$ on a training dataset $\bD$. We use a slight variation of this framework introduced in \cite{nachmani_voice_2020}: the model $f_\theta$ is composed of $R$ blocks $f_\theta^{(1)},\ldots,f_\theta^{(R)}$ which iteratively refine the estimate $\hat{y}$ and provide a sequence $\hat{y}^{(1)}, \ldots, \hat{y}^{(R)}\in\bR^{l\times n}$. Accordingly, $f_\theta$ is optimized using the cumulative expected loss $\bE_{(x,y)\sim\bD}\left[\sum_{r=1}^R\cL(y, \hat{y}^{(r)})\right]$.

\subsection{Separation loss}

The separation loss $\cL$ relies on a pairwise comparison metric $\ell$ tailored for audio signals. We follow \cite{nachmani_voice_2020} and consider SI-SDR \cite{le2019sdr} (referred to as SI-SNR in their work) as the underlying metric:
\begin{equation} \label{eq:sisnr}
    \ell(y, \hat{y}) = 10\log_{\mathrm{10}}\frac{\langle y, \hat{y} \rangle^2}{\|y\|^2\| \hat{y}\|^2 - \langle y, \hat{y} \rangle^2} \;.
\end{equation}

Let $\Sigma$ denote the set of permutations on $\bn$, $\Pi\subset\bR^{n\times n}$ the set of doubly stochastic matrices, and $H$ the entropy. 
In this work, we consider as baselines the standard PIT and SinkPIT training objectives, which can be formulated as follows.
\begin{align}
    \cL_{\mathrm{PIT}}(y, \hat{y}) &=  \min_{\sigma\in\Sigma}\frac{1}{n}\sum_{i=1}^n \ell(y_i, \hat{y}_{\sigma(i)}) \label{eq:pit_loss} \\
    \cL_{\mathrm{SinkPIT}}(y, \hat{y}) &= \min_{\pi\in\Pi}\sum_{i=1}^n\sum_{j=1}^n \pi_{i,j}\ell(y_i, \hat{y}_j) - \varepsilon H(\pi) \label{eq:emd_loss}
\end{align}
It can be proved that both objectives provide the optimal prediction--target matching, when $\varepsilon$ vanishes to 0 in the case of SinkPIT \cite{tachibana_towards_nodate}.

\subsection{Multiple Choice Learning}

MCL has been originally proposed for ambiguous tasks with a single target $y\in\bR^l$ \cite{guzman2012multiple}. This framework trains a neural network $f_\theta$ to provide a small set of $n$ plausible hypotheses $\hat{y}_1,\ldots,\hat{y}_n\in\bR^l$ using a competitive training scheme that promotes the specialization of the hypotheses in distinct regions of the prediction space $\bR^l$. It can be seen as a gradient-descent version of the popular K-means clustering algorithm \cite{lloyd1982least}. Specifically, at each step of the algorithm, the target $y$ is assigned to the closest hypothesis $\hat{y}_i$, and this winning hypothesis is updated by taking a gradient step on the loss $\ell(y, \hat{y}_i)$. The resulting objective is called the Winner Takes All (WTA) loss.

In speech separation, several targets $y_1,\ldots,y_n\in\bR^l$ must be tracked by the hypotheses. To account for this change, the authors of \cite{perera2024annealed} propose to optimize the average target-wise WTA loss. 
\begin{equation}\label{eq:mcl_loss}
    \cL_{\mathrm{MCL}}(y, \hat{y}) = \frac{1}{n} \sum_{i=1}^n \min_{j\in\bn}\ell(y_i, \hat{y}_j) 
\end{equation}
No mechanism ensures that all the hypotheses $\hat{y}_i$ are selected, an issue known as collapse \cite{rupprecht2017learning}. In particular, MCL is not guaranteed to find the optimal prediction--target matching. This is unlike PIT and SinkPIT, which therefore act as upper bound references in our experiments. Nonetheless, we demonstrate empirically that this problem is not encountered in practice (see Section \ref{sec:mcl_pit_comparison}).

\subsection{AUC-SDR}\label{sec:auc_sdr}

\begin{figure}
    \centering
    \includegraphics[width=\linewidth]{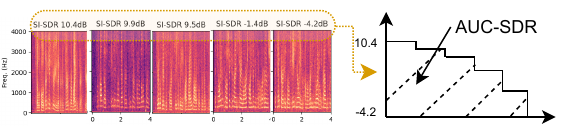}
    \caption{Schematic representation of AUC-SDR.}
    \label{fig:auc-sdr}
\end{figure}

Optimal permutation SI-SDR, which is the customary metric to evaluate speech separation systems, is defined as the average SI-SDR of the $n$ prediction--target pairs. It does not reflect the distribution of SI-SDR scores among these pairs. Therefore, it may fail to penalize scenarios where a few speakers are very well separated at the expense of the others. These scenarios are rare in the few-speaker setting, where the separation performance is uniform across pairs. However, they are common as soon as the number of speakers $n$ increases \cite{dovrat_many-speakers_2021}. 

In order to capture this behavior, we introduce a new metric~(AUC-SDR) which measures the consistency of separation performance across prediction--target pairs (see Figure \ref{fig:auc-sdr}). It is computed as follows. First, we find the optimal prediction--target pairs, then evaluate the SI-SDR score of each pair, and sort these scores in decreasing order $s_1\geq\ldots\geq s_n$. This distribution of scores is normalized to the unit interval $[0,1]$ by mapping the highest score $s_1$ to $1$ and the score lower bound $\mathrm{min}(0,s_n)$ to $0$. Then, AUC-SDR is defined as the empirical mean of these normalized scores, which can also be seen as the area under their curve. This gives a score inside $[0,1]$, where a value close to $1$ indicates perfect consistency, and a value close to $0$ indicates that only few speakers have been correctly reconstructed. 

\section{Experimental Setup}\label{sec:setup}

\subsection{Data}

We report performances on the customary Wall Street Journal (WSJ0-mix) datasets, which consists of synthetic mixtures of clean read speech with 2 to 5 speakers \cite{hershey2016deep}. Each dataset provides 20,000 mixtures for training, 5,000 for validation and 3,000 for test. We additionally report results on the more challenging LibriMix datasets \cite{cosentino2020librimix} with 10 and 20 speakers, which have been introduced by \cite{tachibana_towards_nodate} and \cite{dovrat_many-speakers_2021}. Each dataset provides 1,000 mixture for validation and testing. Additionally, the 10 ans 20 speakers datasets provide, respectively 10,100 mixtures  and 5000 mixtures for training. Each audio recordings is sampled at 8kHz and has a 10-s duration, from which we extract 4s random crops. We use a batch size of $B$ during training, and $1$ during validation and testing.

\begin{table}[t]
\caption{Swave model parameters on WSJ0-mix and LibriMix datasets}
\begin{center}
\resizebox{\columnwidth}{!}{\begin{tabular}{cccc}
\hline
\multirow{ 2}{*}{\textbf{Parameters}} & \multirow{ 2}{*}{\textbf{Symbols}}  & \multicolumn{2}{c}{\textbf{Values}} \\
\cline{3-4}
 &  & \textbf{WSJ0-mix} & \textbf{LibriMix} \\
\hline
Number of features & N & 128 & 256 \\
\hline
Encoder's kernel size & L & 8 & 16 \\
\hline
Number of hidden units in the LSTMs & H & 128 & 256 \\
\hline
Number of double MulCat blocks & R & 6 & 7 \\
\hline
Batch size & B & 4 & max \\
\hline
Learning rate & $\lambda$ & 5e-4 & 1e-3 \\
\hline
Device &  & V100 & A100 \\
\hline
\end{tabular}}
\label{tab:parameters}
\end{center}
\end{table}

\subsection{Model}

Many efficient neural network architectures have been devised to tackle speech separation through recent years \cite{luo2019conv, luo_dual-path_2020, subakan_attention_2021, lutati_sepit_2023, zhao2023mossformer, wang2023tf,  lee2024boosting, perera2024annealed}. However, most of these models are very large (up to 26M parameters with SepFormer \cite{subakan_attention_2021}) and have been designed for the few-speakers setting (from 2 to 5 speakers). In contrast, the authors of \cite{nachmani_voice_2020} proposed Swave, a 7.5M-parameter model with performances close to the state of the art, which can separate as many as 20 speakers given slight modifications \cite{dovrat_many-speakers_2021}. We base our experiments on this work.

This model is composed of a 1d convolutional encoder with kernel size $L$, $R$ MulCat 
blocks and a decoder which performs an overlap and add operation. A MulCat block consists of two separate bidirectional LSTMs with $N$ features and $H$ hidden units, whose outputs are multiplied element-wise and concatenated to generate the final output. Following the original implementation, we use different configurations for WSJ0-mix and LibriMix. Their exact values are specified in Table \ref{tab:parameters}. Swave has 7.5M parameters for the WSJ0-mix datasets, and 36.5M parameters for the LibriMix dataset. All models are trained for 40 epochs, except on the LibriMix dataset with 10 speakers, for which we use 20 epochs instead.  We use the Adam optimizer with learning rate $\lambda$. We use Nvidia V100 and A100 GPUs.

\subsection{Metric}

All metrics are computed on the test sets. We report the optimal permutation SI-SDR as a performance metric to compare trained networks, as done in \cite{nachmani_voice_2020,dovrat_many-speakers_2021}. In order to ensure fair comparison, we train MCL, PIT and SinkPIT using the same experimental settings, and report performances in Table \ref{tab:score}. For PIT, all variants match or outperform the scores reported in the original papers (less than 1dB of gap in the worst case), and the slight observed discrepancies are due to differences in training time (the authors use a longer training of 100 epoch).

\section{Results}\label{sec:results}

\subsection{Performance}\label{sec:mcl_pit_comparison}

Comparing lines 1 and 2 of Table \ref{tab:score}, we observe that MCL performs on par with the topline approach PIT, in the few-speaker and the many-speaker settings alike. This establishes MCL as a viable alternative to PIT for speech separation. The same holds for SinkPIT, so that the three methods perform equivalently in terms of separation performance.

We can also highlight that these results were obtained with a fraction of the number of training epochs reported in the original papers, which suggests that Swave is a very efficient and versatile model.

\subsection{Time complexity}

\begin{figure}[htbp]
\centering{\includegraphics[width=\linewidth]{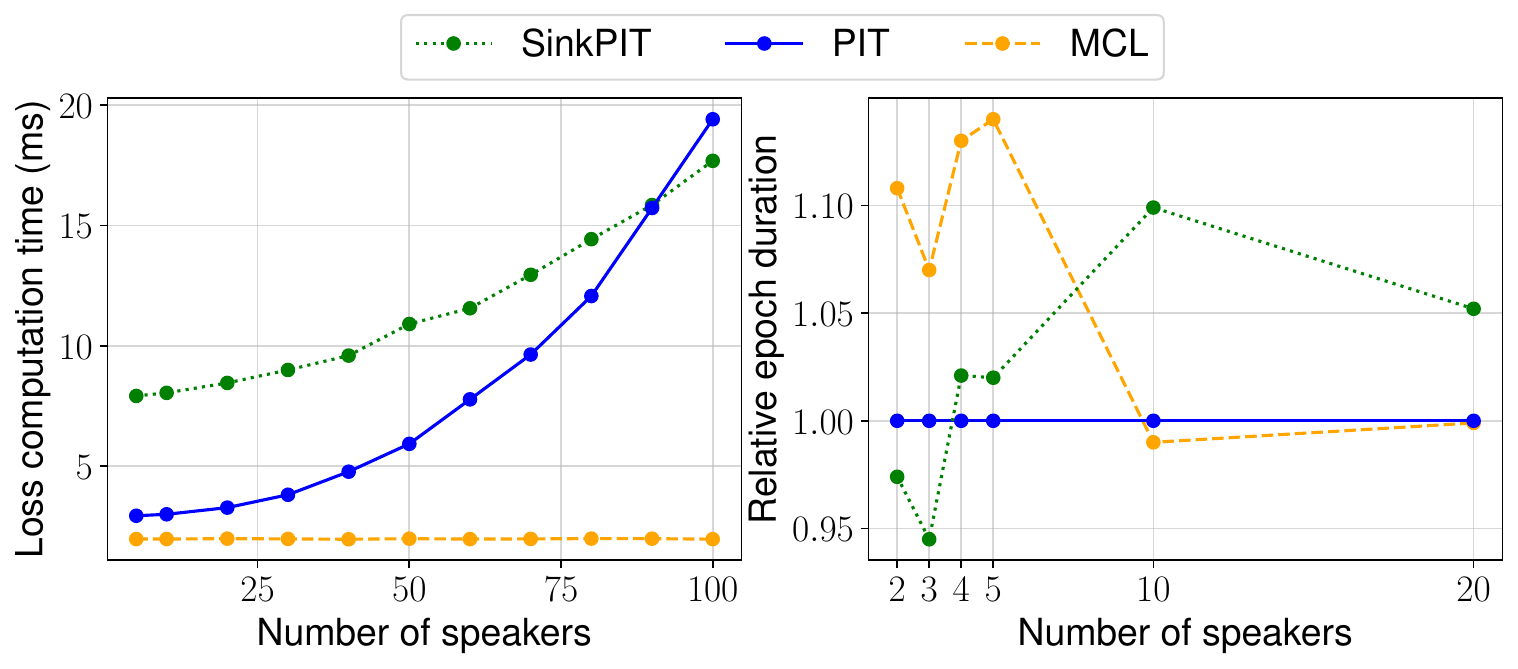}}
\caption{\textbf{Time complexity of MCL, PIT and SinkPIT.} On the left, we show the average computation time per sample of MCL (orange dashed line), PIT (blue solid line) and SinkPIT (green dotted line) separation losses, as a function of the number of speakers. On the right, we display the relative training time over one epoch of MCL and SinkPIT, computed for the WSJ0-mix datasets (2 to 5 speakers) and LibriMix datasets (10 and 20 speakers), as a function of the number of speakers. For each dataset, the training time of PIT serves as the reference and is represented by a value of 1 (blue solid line).}
\label{fig:time_complexity}
\end{figure}

We also compare the three approaches in terms of training complexity, and present the results in Figure \ref{fig:time_complexity}. First, we can notice that MCL and SinkPIT have an efficient algorithmic complexity of $\cO(n^2)$ while PIT is $\cO(n^3)$. We observe experimentally that the average computation time per sample of MCL, PIT and SinkPIT separation losses indeed reflect this theoretical analysis when the number of speakers grows (left panel of Figure \ref{fig:time_complexity}). This strengthens the previous observation, and suggests that MCL is a sound alternative to PIT.

However, this difference in computation time is not reflected in the average epoch duration (right panel of Figure \ref{fig:time_complexity}). Indeed, for $n\leq20$ speakers, the computation time gap between PIT, MCL and SinkPIT separation losses becomes negligible compared to other factors in the training pipeline (e.g., data loading). In this regime, we observe only marginal differences in epoch duration between the three methods.  Nonetheless, optimal permutation search becomes the main bottleneck for the training time as soon as $n$ approaches 100. Such an extreme scenario is unlikely for speech separation, but it may occur in different context (environmental sound separation \cite{kavalerov2019universal}, neural activity analysis \cite{zheng2001signal}).

\subsection{Separation consistency}

We compute AUC-SDR (as described in Section \ref{sec:auc_sdr}) for all compared methods and report the results in Table \ref{tab:AUC_SDR}. First, we observe that the separation consistency decreases as the number of speakers grows. The separation performance seems to stabilize to 0.5 when there are many speakers, which suggests that a majority of speakers are well separated even in this challenging setting. 

Second, we observe that for each dataset, there are only marginal differences in the separation consistency of PIT, MCL and SinkPIT. This further demonstrates the  viability of MCL for speech separation, and suggests that it manages to find the optimal prediction--target mapping. 

Third, the AUC-SDR scores obtained by PIT on the 10-speaker and 20-speaker datasets are far from perfect consistency. This suggests that optimal matching based losses are not sufficient to ensure high separation consistency. The design of training objectives that enforce separation consistency is left to future work.

\begin{table}[htbp]
\caption{Evaluation using SI-SDR [dB]}
\begin{center}
\resizebox{\columnwidth}{!}{\begin{tabular}{|c|cccc|cc|}
\hline
&\multicolumn{4}{c|}{WSJ0-mix} &\multicolumn{2}{c|}{LibriMix}\\
\cline{2-7} 
Loss & 2 spks & 3 spks & 4 spks & 5 spks & 10 spks$^*$ & 20 spks\\
\hline
\hline
PIT &     19.96 &   16.13 &   12.00 &   10.14 &      3.17 & 2.98 \\
\hline
MCL   &   19.74 &   16.01 &   11.85 &    9.92 &     3.85 & 2.96 \\
\hline
SinkPIT   &   19.63 &   16.00 &   12.12 &   10.22 &     3.50 & 3.10\\
\hline
\end{tabular}}
\footnotesize{$^*$ Scores for this dataset are computed after 20 epochs.}
\label{tab:score}
\end{center}
\end{table}

\begin{table}[htbp]
\caption{Evaluation using AUC-SDR}
\begin{center}
\resizebox{\columnwidth}{!}{\begin{tabular}{|c|cccc|cc|}
\hline
&\multicolumn{4}{c|}{WSJ0-mix} &\multicolumn{2}{c|}{LibriMix}\\
\cline{2-7} 
Loss & 2 spks & 3 spks & 4 spks & 5 spks & 10 spks & 20 spks\\
\hline
\hline
PIT &  0.93  &  0.81  & 0.65 & 0.57 & 0.50 & 0.48 \\
\hline
MCL   & 0.93 & 0.81 & 0.64 & 0.57 & 0.54 & 0.49 \\
\hline
SinkPIT   & 0.93 &  0.80 & 0.65 & 0.57 & 0.52 & 0.49 \\
\hline
\end{tabular}}
\label{tab:AUC_SDR}
\end{center}
\end{table}

\subsection{MCL collapse}

MCL is subject to collapse issues \cite{rupprecht2017learning}. In speech separation, this corresponds to the situation where only a subset of the predictions correctly estimate the individual speech targets, and the other predictions are inaccurate.
The authors of \cite{perera2024annealed} suggest to use annealing in order to mitigate collapse. Although this device does not seem necessary to match the performances of PIT, it may further improve separation consistency.

\subsection{Discussion} 

The results discussed above establish that MCL is a strong substitute to PIT. This is an interesting result, because MCL has many natural extensions that makes it fit for more challenging settings. First, MCL can be extended to scenarios with very large number of sources, at minimal increase of the computational cost. Second, one of the key challenges in source separation is to handle settings with a variable number of speakers \cite{ochieng2023deep}. MCL provides an elegant way to tackle this issue by using scoring heads, similarly to \cite{letzelter2024resilient}. Third, by observing the strong specialization capabilities of MCL \cite{rupprecht2017learning}, we can infer that this framework can be used in an unsupervised fashion, as long as it is trained on a dataset with few simultaneously active speakers.

\section{Conclusion}\label{sec:conclusion}

In this article, we show that MCL can replace PIT, the standard objective used to train speech separation models. Although MCL is not guaranteed to find an optimal solution to the permutation problem tackled by PIT, we demonstrate experimentally that MCL is on par with PIT, in terms of separation performance, training complexity, and separation consistency (as measured using our newly introduced AUC-SDR metric).
This opens an interesting venue for further research, because MCL has many advantages over PIT: more efficient asymptotic time complexity as the number of speakers grows large, ability to handle a variable number of speakers, and adaptability to the more challenging unsupervised setting. A more careful analysis of these settings is left to future work.

\section{Acknowledgement}

The material contained in this document is based upon work funded by the Agence National de la Recherche en Intelligence Artificielle (PhD program in AI) and Hi! PARIS through its PhD funding program. This work was performed using HPC resources from GENCI–IDRIS (Grant 2021-AD011013406R1).

\bibliographystyle{IEEEtran}
\bibliography{bibliography}

\end{document}